\PassOptionsToPackage{square}{natbib}
\PassOptionsToPackage{sort&compress}{natbib}
\PassOptionsToPackage{numbers}{natbib}

\documentclass[preprint,12pt]{elsart}

\usepackage{graphicx}
\graphicspath{ {figure_mmgarticle/} }

\usepackage{amssymb}

\usepackage{lineno}
\usepackage{upgreek}
\usepackage[square,sort&compress,numbers]{natbib}

\journal{Physics Letters B}

\newcommand{\affuni}[2]{Dipartimento di Fisica dell'Universit\`a #1, #2, Italy.}
\newcommand{\affinfn}[2]{INFN Sezione di #1, #2, Italy.}
\newcommand{\hu}{$e^+ e^- \rightarrow U h^\prime$}

\newcommand{\mmg}{$e^+ e^- \rightarrow \mu^+ \mu^- \gamma$}

\begin{document}
\begin{frontmatter}
\title{Search for dark Higgsstrahlung in $e^+ e^- \rightarrow \mu^+ \mu^-$ 
and missing energy 
events with the KLOE experiment}
\collab{The KLOE-2 Collaboration}
\author[Frascati]{D.~Babusci},
\author[Frascati]{G.~Bencivenni},
\author[Frascati]{C.~Bloise},
\author[Frascati]{F.~Bossi},
\author[INFNRoma3]{P.~Branchini},
\author[Roma3,INFNRoma3]{A.~Budano},
\author[Uppsala]{L.~Caldeira~Balkest\aa hl},
\author[Roma3,INFNRoma3]{F.~Ceradini},
\author[Frascati]{P.~Ciambrone},
\author[Messina,INFNCatania]{F.~Curciarello},
\author[Cracow]{E.~Czerwi\'nski},
\author[Frascati]{E.~Dan\`e},
\author[INFNRoma3]{V.~De~Leo},
\author[Frascati]{E.~De~Lucia},
\author[Frascati]{A.~De~Santis},
\author[Frascati]{P.~De~Simone},
\author[Roma3,INFNRoma3]{A.~Di~Cicco},
\author[Roma1,INFNRoma1]{A.~Di~Domenico},
\author[INFNRoma2]{R.~Di~Salvo},
\author[Frascati]{D.~Domenici},
\author[Roma2,INFNRoma2]{A.~Fantini},
\author[Frascati]{G.~Felici},
\author[ENEACasaccia,INFNRoma1]{S.~Fiore},
\author[Cracow]{A.~Gajos},
\author[Roma1,INFNRoma1]{P.~Gauzzi},
\author[Messina,INFNCatania]{G.~Giardina},
\author[Frascati]{S.~Giovannella},
\author[INFNRoma3]{E.~Graziani}\ead{enrico.graziani@roma3.infn.it},
\author[Frascati]{F.~Happacher},
\author[Uppsala]{L.~Heijkenskj\"old},
\author[Uppsala]{W.~Ikegami Andersson},
\author[Uppsala]{T.~Johansson},
\author[Cracow]{D.~Kami\'nska},
\author[Warsaw]{W.~Krzemien},
\author[Uppsala]{A.~Kupsc},
\author[Roma3,INFNRoma3]{S.~Loffredo},
\author[Messina,INFNCatania]{G.~Mandaglio},
\author[Frascati,Marconi]{M.~Martini},
\author[Frascati]{M.~Mascolo},
\author[Roma2,INFNRoma2]{R.~Messi},
\author[Frascati]{S.~Miscetti},
\author[Frascati]{G.~Morello},
\author[INFNRoma2]{D.~Moricciani},
\author[Cracow]{P.~Moskal},
\author[INFNRoma3,LIP]{F.~Nguyen}\ead{federico.nguyen@cern.ch},
\author[Frascati]{A.~Palladino},
\author[INFNRoma3]{A.~Passeri},
\author[Energetica,Frascati]{V.~Patera},
\author[INFNBari]{A.~Ranieri},
\author[Frascati]{P.~Santangelo},
\author[Frascati]{I.~Sarra},
\author[Calabria,INFNCalabria]{M.~Schioppa},
\author[Frascati]{M.~Silarski},
\author[INFNRoma3]{L.~Tortora},
\author[Frascati]{G.~Venanzoni},
\author[Warsaw]{W.~Wi\'slicki},
\author[Uppsala]{M.~Wolke}
\address[INFNBari]{\affinfn{Bari}{Bari}}
\address[INFNCatania]{\affinfn{Catania}{Catania}}
\address[Calabria]{\affuni{della Calabria}{Cosenza}}
\address[INFNCalabria]{INFN Gruppo collegato di Cosenza, Cosenza, Italy.}
\address[Cracow]{Institute of Physics, Jagiellonian University, Cracow, Poland.}
\address[Frascati]{Laboratori Nazionali di Frascati dell'INFN, Frascati, Italy.}
\address[Messina]{Dipartimento di Fisica e Scienze della Terra dell'Universit\`a di Messina, Messina, Italy.}\address[Moscow]{Institute for Theoretical and Experimental Physics (ITEP), Moscow, Russia.}
\address[Energetica]{Dipartimento di Scienze di Base ed Applicate per l'Ingegneria dell'Universit\`a 
``Sapienza'', Roma, Italy.}
\address[Marconi]{Dipartimento di Scienze e Tecnologie applicate, Universit\`a ``Guglielmo Marconi", Roma, Italy.}
\address[Roma1]{\affuni{``Sapienza''}{Roma}}
\address[INFNRoma1]{\affinfn{Roma}{Roma}}
\address[Roma2]{\affuni{``Tor Vergata''}{Roma}}
\address[INFNRoma2]{\affinfn{Roma Tor Vergata}{Roma}}
\address[Roma3]{Dipartimento di Matematica e Fisica dell'Universit\`a 
``Roma Tre'', Roma, Italy.}
\address[INFNRoma3]{\affinfn{Roma Tre}{Roma}}
\address[ENEACasaccia]{ENEA UTTMAT-IRR, Casaccia R.C., Roma, Italy}
\address[Uppsala]{Department of Physics and Astronomy, Uppsala University, Uppsala, Sweden.}
\address[Warsaw]{National Centre for Nuclear Research, Warsaw, Poland.}
\address[LIP]{Present Address: Laborat\'orio de Instrumenta\c{c}\~{a}o e F\'isica Experimental de Part\'iculas,
Lisbon, Portugal.}

\begin{abstract}
We searched for evidence of a Higgsstrahlung process in a secluded sector, 
leading to a final state with a dark photon $U$ and a dark Higgs boson 
 $h^\prime$, with the KLOE detector at DA$\Phi$NE.
We investigated the case of $h^\prime$ lighter than $U$, 
with $U$ decaying into a muon pair and $h^\prime$ producing
a missing energy signature. 
We found no evidence of the process and set upper limits 
to its parameters in the range
$2 m_\mu < m_U <$ 1000 MeV, $m_{h^\prime} < m_U$.  
\end{abstract}


\end{frontmatter}

\clearpage

\section{Introduction}
\label{Introduction}
Astrophysical data reveal in a more and more convincing way that our
knowledge of the Universe is limited to about 4-5\% of the total 
matter-energy content: this is generally interpreted as an evidence 
of the existence of dark matter and dark energy components.  
In recent years, several astrophysical observations have failed 
to find a common interpretation in terms of standard astrophysical or particle 
physics 
sources \cite{integral,pamela,ams,atic,fermi,hess1,hess2,dama1,dama2,cogent}. 
Although there are alternative explanations for some of these
results, they could all be explained with the existence of a dark
matter weakly interacting massive particle, WIMP, belonging to a
secluded gauge sector under which the Standard Model (SM) particles
are uncharged 
\cite{dieci,undici,dodici,tredici,quattordici,
quindici,sedici,diciassette,diciotto,diciannove}. 
In a minimal model, a new abelian $U(1)_S$ gauge field is introduced, 
the U boson or dark photon,
with mass near the GeV scale, coupled  to the
SM only through its kinetic mixing with the SM hypercharge 
field. 
The kinetic mixing parameter $\epsilon$  is expected to be of 
the order $10^{-4}-10^{-2}$ 
\cite{undici,dodici,tredici,
quattordici,quindici,sedici,diciassette,diciotto,diciannove,venti}, 
so that observable effects can be detected
at $e^+e^-$ colliders 
\cite{ventuno,venti,ventidue,
ventitre,ventiquattro} 
or at fixed target experiments working in the GeV region 
\cite{venticinque,ventisei,ventisette,ventotto}. 
The existence of the U boson, through its mixing with the ordinary photon,
can also accomodate the observed discrepancy in the measured 
muon anomalous magnetic
moment $a_\mu$ with respect to the 
SM prediction \cite{gmenodue}.
Several searches of the
$U$ boson have been performed in recent years with negative results,
setting upper limits to $\epsilon$: A1 \cite{a1_1,a1_2}, APEX \cite{apex}, 
WASA \cite{wasa}, HADES \cite{hades}, KLOE \cite{kloe1,kloe2}, 
BaBar \cite{babar2}.

Since the U boson needs to be massive, one can implement, in close analogy 
with the SM, a spontaneous breaking mechanism of 
the $U(1)_S$ symmetry, thus  introducing a Higgs-like
particle, $ h^\prime$ or dark Higgs, whose mass hierarchy with the dark photon
is not constrained by the theory \cite{ventuno}.

The U boson can be produced at $e^+e^-$ colliders via different
processes: $e^+e^- \rightarrow U \gamma$ , $e^+e^- \rightarrow U h^\prime$ 
(dark Higgsstrahlung) 
and in decays of vector particles to pseudoscalars. 
In this work the Higgsstrahlung
process $e^+e^- \rightarrow U h^\prime$  is studied, using data collected 
by the KLOE 
experiment at the $e^+e^-$ collider DA$\Phi$NE at the Frascati
laboratory, both at a center of
mass energy of $\sim$ 1019 MeV, the mass of the $\phi$ meson (on-peak sample), and  
at a center of mass energy of $\sim$ 1000 MeV (off-peak sample). 
The process  $e^+e^- \rightarrow U h^\prime$, with $U$ decaying
into lepton or hadron pairs, is an interesting 
reaction to be studied at an  $e^+e^-$ collider, being less suppressed,
in terms of the mixing parameter, 
than the other final states listed above.  
There are two very different scenarios 
depending on the masses of the dark photon $m_U$ and of the dark 
Higgs boson  $m_{h^\prime}$. 
For $m_{h^\prime}$ larger than 2$m_U$, the dark Higgs 
boson would decay dominantly and promptly to a U boson pair, thus giving
rise to a six charged particle final state (the scenario with  $m_{h^\prime}$ 
larger than $m_U$ but smaller than $2 m_U$ is similar, with one dark photon 
off shell):
this case was recently investigated by the 
BaBar \cite{babar} and Belle \cite{belle} experiments. 
On the other side, Higgs bosons lighter
than the dark photon would have, in most of the parameter space region,
such a large lifetime to escape detection,
showing up as a missing energy signature. 
We confined the search only to the latter case, $m_{h^\prime} < m_U$,
the so called ``invisible'' dark Higgs scenario. 

The lifetime of the dark Higgs boson depends on the kinetic mixing parameter
$\epsilon$, the boson masses $m_{h^\prime}$ and $m_U$ and the
dark coupling constant $\alpha_D$ \cite{ventuno}. 
For boson masses of the order of 100 MeV and 
$\alpha_D = \alpha_{em}$, the dark Higgs boson lifetime would be $\sim 5 \mu s$ 
for $\epsilon \sim 10^{-3}$, corresponding, for the energy
range explored in this analysis, to a decay 
length of $\sim $ 100 m. The dark Higgs boson would be thus invisible up 
to $\epsilon \sim 10^{-2} \div 10^{-1}$, depending on the $h^\prime$ mass.

In this work the search is limited to the decay of the U boson 
in a muon pair: the final state 
signature is then a pair of opposite charge muons plus missing energy. 
The measurement is thus performed in the range 
$2 m_\mu < m_U <$ 1000 MeV with the constraint $m_{h^\prime} < m_U$. 

The production cross section of the dark Higgsstrahlung process is 
proportional to the product $\alpha_D \times \epsilon^2$ 
and depends on the boson masses \cite{ventuno}.
Values as high as hundreds of fb are reachable in this model. 
Compared to the B-factory case \cite{babar, belle}, KLOE benefits of the 
1/s factor and of the resonance-like behaviour expected for the production 
cross section \cite{ventuno}.   
The branching ratio of the U boson into muon pairs
is predicted to be just 
below the 50\% level for masses slightly above the kinematical threshold 
$m_U=2 m_\mu$, then to decrease 
up to a minimum around 5\%, for masses corresponding to the $\rho$ resonance 
(due to the concurrent decay into hadrons), 
and then to increase to 
$\sim 30\div40 \%$ up to $m_U \simeq$ 1 GeV \cite{ventuno}.

\section{The KLOE Detector}
\label{KLOE}

DA$\Phi$NE, the Frascati $\upphi$-factory, is an $\mathrm{e}^+\mathrm{e}^-$  
collider working at the center of mass energy, 
$\sqrt{s}\sim m_\upphi=1.0195$ GeV \cite{dafne}.    
Positron and electron beams collide at an angle of $\pi-$25 mrad, 
producing $\mathrm{\upphi}$ mesons nearly at rest. 
The KLOE detector is made up of a large cylindrical drift 
chamber (DC)~\cite{trentaquattro}, surrounded by a lead scintillating fiber 
electromagnetic calorimeter (EMC)~\cite{trentacinque}. 
A superconducting coil around  the EMC provides a 0.52 T magnetic 
field along the axis of the colliding beams.

The EMC consists of barrel and end-cap modules covering 98\% of the 
solid angle. The calorimeter modules are 
segmented into five layers in depth and 
read out at both ends by 4880 
photomultipliers. Energy and time resolutions 
are $ \sigma_E /E=~0.057 /\sqrt{E(\mathrm{GeV})} $ 
and $ \sigma_t =57 \, \mathrm{ps}~ /\sqrt{E(\mathrm{GeV})}~\oplus 100\, \mathrm{ps} $, respectively.
The drift chamber, with only stereo sense wires, $4$ m in diameter 
and $ 3.3$ m long, has a mechanical structure in carbon-fiber and 
operates with a low-mass gas mixture (90\% helium, 10\% isobutane). 
The spatial resolutions are  $\sigma_{xy} \sim 150\, \mathrm{\upmu m}$ 
and  $\sigma_z \sim2$ mm. 
The momentum resolution for large angle tracks 
is $\sigma_{p_\perp} / p_{\perp} \approx 0.4\% $. 
The trigger~\cite{trentasei} uses both 
EMC and DC information. 
Data are then analysed by an event classification filter 
\cite{trentasette} , which selects
and streams various categories of events in different output files.

\section{Event selection}
The analysis of the process $e^+e^- \rightarrow U h^\prime$ , 
$U \rightarrow \mu^+ \mu^-$, $h^\prime$ invisible (\hu\,in the following), 
has been performed on a data
sample of 1.65 fb$^{-1}$ collected 
at a center of mass energy of $\sim$ 1019 MeV, corresponding to
the mass of the $\phi$ meson (on-peak sample in the following), 
and on a data sample of 0.206 fb$^{-1}$ at a 
center of mass energy of $\sim$ 1000 MeV (off-peak sample in the following), 
well below the $\phi$ resonance.
\begin{figure}[htp!]
    \begin{center}
        {\includegraphics[width=9cm]{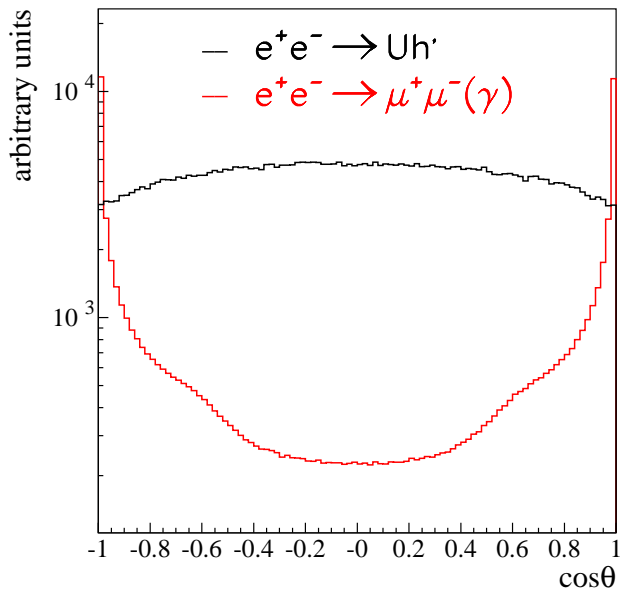}}
        \caption{Distribution of the polar angle of the muon pair momentum
for the signal \hu \, (black line) 
and for \mmg \, (red line). 
Here the two processes are not normalised and are shown
only in order to compare the shapes of the distributions. All the generated 
samples at various $m_U$ and $m_{h^\prime}$ are included in the signal 
distribution.}
\label{figcos}
    \end{center}
\end{figure}

The Monte Carlo simulation of the signal process \hu\,has 
been produced using an ad hoc generator interfaced
with the standard KLOE simulation program \cite{trentasette}. 
Signal  samples have been generated for various pairs of  
$m_{h^\prime}$ - $m_U$ values along a grid
with steps of $\sim$ 30 MeV  
to cover all the allowed kinematic region. The invariant 
mass resolution varies between 0.5 and 2 MeV 
for the muon pair ($M_{\mu \mu}$), 
and between 3 and 17 MeV  
for the event missing mass ($M_{miss}$).
The signal process signature would thus be the appearance of a
sharp peak in the bidimensional distribution 
$M_{\mu \mu}$ - $M_{miss}$.   
Moreover, the distribution of the polar angle direction of the muon pair 
momentum, $\theta$,
contrarily to most of the dominant background
processes, is expected to prefer large angles.
The differential cross section has two 
dominant terms proportional to $\sin\theta$ 
and $\sin^3\theta$ \cite{ventuno},  
with relative weights smoothly dependent on the boson masses.
This 
angular distribution
allows to reject most of the background of QED processes with a 
simple geometrical 
selection and 
implies that the missing momentum direction preferably points to a very
well equipped region of the KLOE detector, where the best efficiency
is achieved.
Fig.\,\ref{figcos} shows the distributions of the 
muon pair polar angle direction
for the signal \hu \, (black line) and the \mmg \, background (red line), where
all the generated samples at various
$m_U$ and $m_{h^\prime}$  masses are included in the signal sample.

As a first step of the analysis, a preselection was performed by requiring:
\begin{itemize}
\item events with only two opposite charge tracks with associated EMC
clusters, 
with polar angles  $|\cos \theta_{1,2}|<0.8$ and momenta 
below 460 MeV, that form a  
reconstructed vertex inside a 
cylinder of 30 cm length, 4 cm radius, centered at the interaction point (IP);
\item the sum of the momenta of the two tracks to be greater than 450 MeV; 
\item the polar angle of the dimuon momentum
to be in the barrel acceptance: $|\cos\theta| < 0.75$;
\item the modulus of the missing momentum to exceed 40 MeV.  
\end{itemize}   

After this selection, mostly aimed at rejecting backgrounds from
QED processes, 
the hermeticity and tightness of the electromagnetic calorimeter 
was used as a veto 
to avoid the presence of photons in the event by requiring  
no prompt EMC clusters 
unassociated to tracks.  
The calorimeter veto inefficiency
as a function of the energy was studied with a sample
of radiative Bhabha scattering events $e^+ e^- \rightarrow e^+ e^- \gamma$ 
and found to range between 10\% at 20 Mev and 0.1\% at about 200 MeV.

The event selection then proceeded by applying a particle identification
(PID) algorithm to the two tracks, based
on the excellent energy and time resolution of the EMC. A set of 
feed-forward neural
networks, organised for different values of track momentum 
and track polar angle,
was trained on simulated Monte Carlo samples
to perform muon to electron discrimination. The neural networks
used five input variables (cluster time, energy to
momentum ratio and three variables related to energy 
depositions in calorimeter layers) and produces one output. 
The PID performances were checked on selected data samples of
$e^+ e^- \rightarrow e^+ e^-$ , $e^+ e^- \rightarrow \mu^+ \mu^-$,
$e^+ e^- \rightarrow \pi^+ \pi^-$: the fraction of events
where both tracks were identified as muons 
was measured to
be 85\% in $e^+ e^- \rightarrow \mu^+ \mu^-$ events, 
10$^{-4}$ in $e^+ e^- \rightarrow e^+ e^-$ events and $\sim$ 50\% 
in $e^+ e^- \rightarrow \pi^+ \pi^-$ events
(showers produced by muons or pions have similar properties at low energies).

After missing energy and PID selections, a large background 
from $\phi \rightarrow K^+ K^-$, $K^\pm \rightarrow \mu^\pm \nu$
events survived in the on-peak sample. 
This happens when both kaons decay semileptonically 
close to the IP. 
Charged kaons have an average decay 
length of $\sim$ 90 cm in KLOE. The reconstructed vertex of the  
muon tracks is thus expected to be displaced from the IP 
and with a bad fit quality.
Cuts on the radial 
and z  
projections of the distance 
between the reconstructed
vertex and the IP and on the $\chi^2$ of the fit 
allowed to reduce by a 
factor $\sim$ 80 the $\phi \rightarrow K^+ K^-$, $K^\pm \rightarrow \mu^\pm \nu$
background, lowering the signal efficiency by $\sim$ 65\%.

Events surviving all the described selections were organized in bidimensional
histograms with the muon pair mass $M_{\mu \mu}$ and the event missing mass 
$M_{miss}$ on the two axes.
The binning was chosen to keep most of the signal inside a single bin.
For $M_{\mu \mu}$ a 5 MeV bin width was enough over all the plane; while 
for $M_{miss}$ a variable binning of 15, 30 and 50 MeV widths was
chosen. According to the simulation, a fraction 
of 90$\div$95\% of the signal was contained in one single bin. 
The signature of the process would thus 
be the appearance
of an excess in a single bin in the  $M_{\mu \mu}$-$M_{miss}$ plane
over the background. 
The signal selection efficiency, estimated from Monte Carlo
on the generated points of the $m_U$-$m_{h^\prime}$ grid, was
found to be between 15\% and 25\%, depending on the masses,
with most frequent values of $\sim$20\%. The efficiency 
for a generic point on the 
$M_{\mu \mu}$-$M_{miss}$ plane was then evaluated by linear interpolation.

\section{Results}
\begin{figure}[htp!]
    \begin{center}
        {\includegraphics[width=6.7cm]{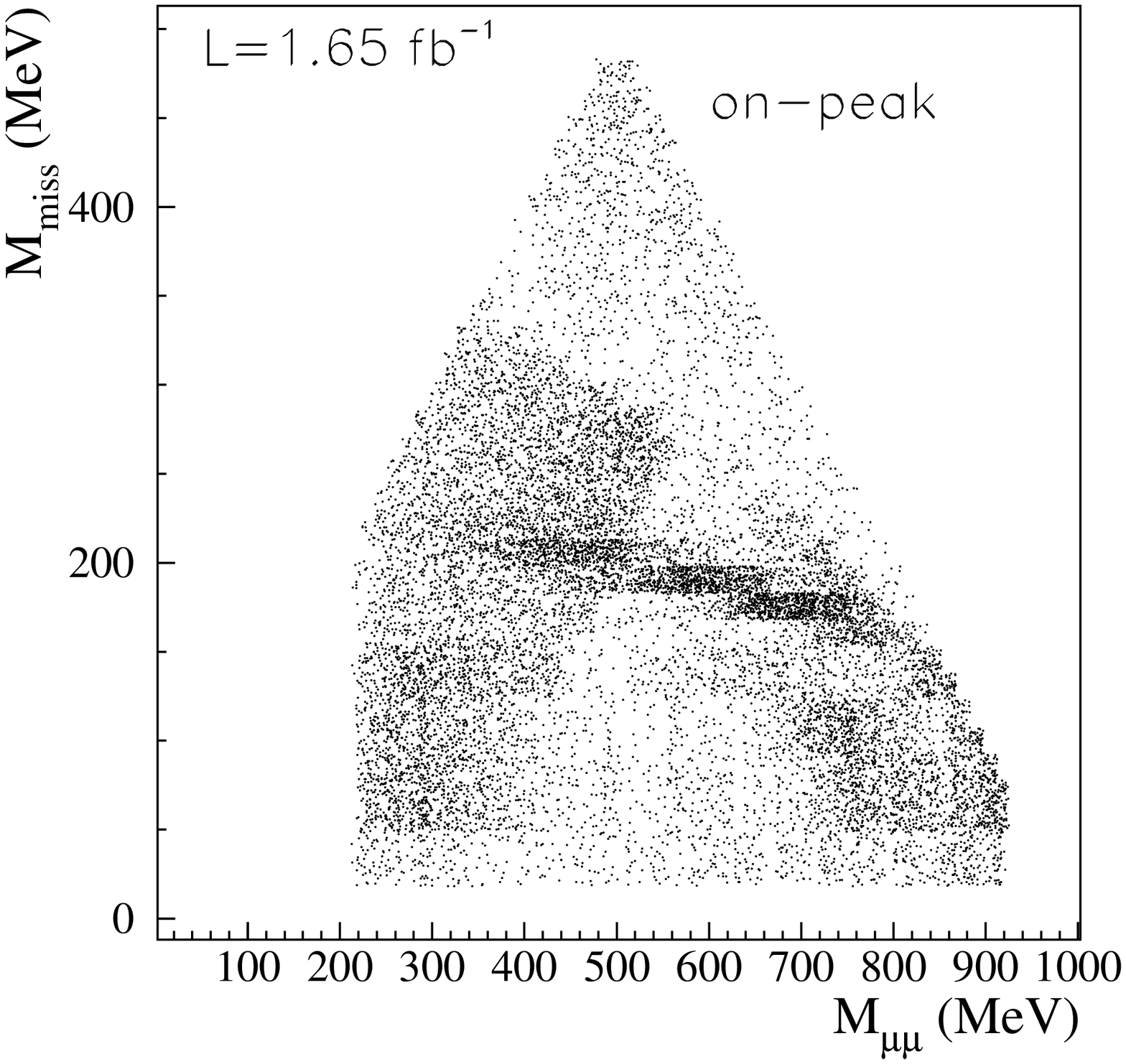}}\hspace{0.3cm}
        {\includegraphics[width=6.7cm]{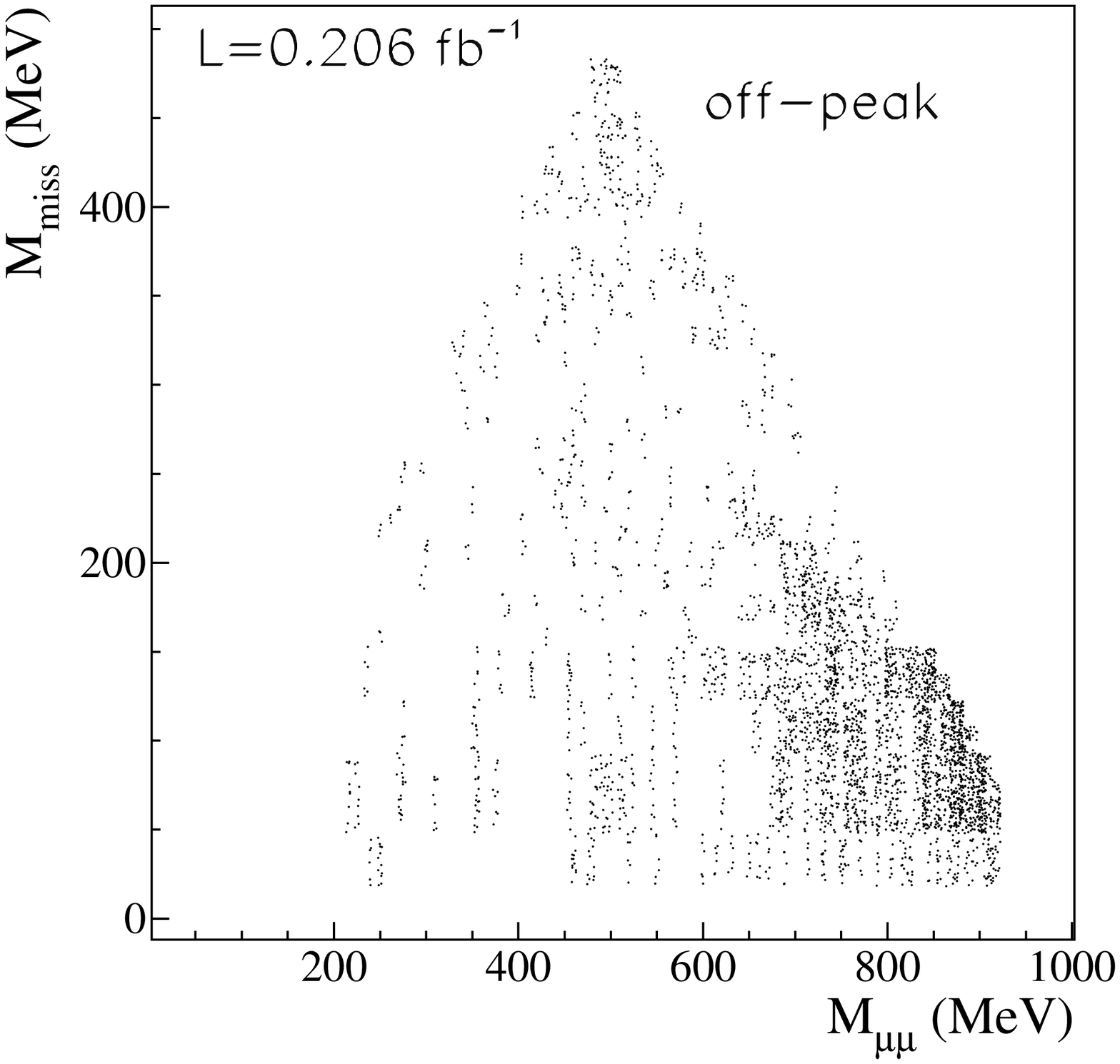}}
        \caption{Results for on-peak sample 
(left plot, 1.65 fb$^{-1}$ integrated luminosity) and off-peak sample 
(right plot, 0.206 fb$^{-1}$ integrated luminosity).}
\label{figdati}
    \end{center}
\end{figure}

After all the described selections, 15278 events survived
in the on-peak sample (fig.\,\ref{figdati}, left plot) 
and 783 in the off-peak sample (fig.\,\ref{figdati}, right plot). 
In the left plot of fig.\,\ref{figdati} (on-peak sample) several 
sources of backgrounds can be distinguished:

\begin{itemize}
\item  $\phi \rightarrow K^+ K^-$, $K^\pm \rightarrow \mu^\pm \nu$ (quadrangular
region at the left of the populated part of the distribution);
\item $\phi \rightarrow \pi^+ \pi^- \pi^0$ (quasi-horizontal band, 
corresponding to events in which both photons from the $\pi^0$ decay 
are undetected), partly
intersecting the $\phi \rightarrow K^+ K^-$, $K^\pm \rightarrow \mu^\pm \nu$ 
region; 
\item $e^+ e^- \rightarrow \mu^+ \mu^-$ and 
$e^+ e^- \rightarrow \pi^+ \pi^-$ events in the continuum 
(diagonal and horizontal bands starting from the 
right-bottom part of the distribution); 
\item $e^+ e^- \rightarrow e^+ e^- \mu^+ \mu^-$ and 
$e^+ e^- \rightarrow e^+ e^- \pi^+ \pi^-$ (photon-photon interactions, top 
triangular part of
the distribution, for $M_{miss} >$ 350 MeV), with $e^\pm$ in the final state
being scattered at very small angles in the beam pipe.
\end{itemize}

In the distribution in the right plot of fig.\,\ref{figdati} (off-peak sample) 
all the backgrounds from the $\phi$ decays are strongly suppressed and only 
those in the continuum remain visible.

\begin{figure}[htp!]
    \begin{center}
        {\includegraphics[scale=0.33]{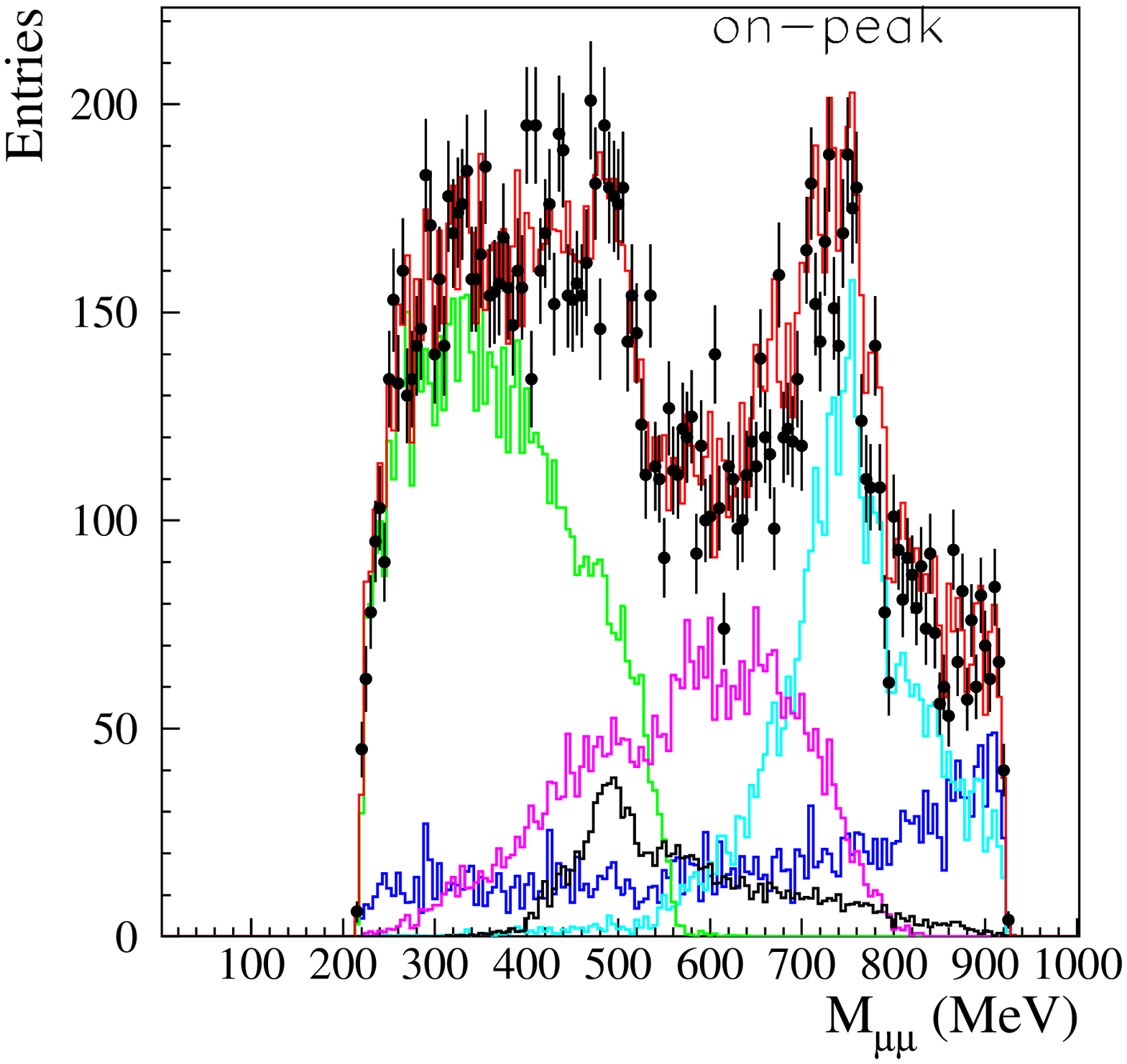}}\hspace{0.5cm}
        {\includegraphics[scale=0.33]{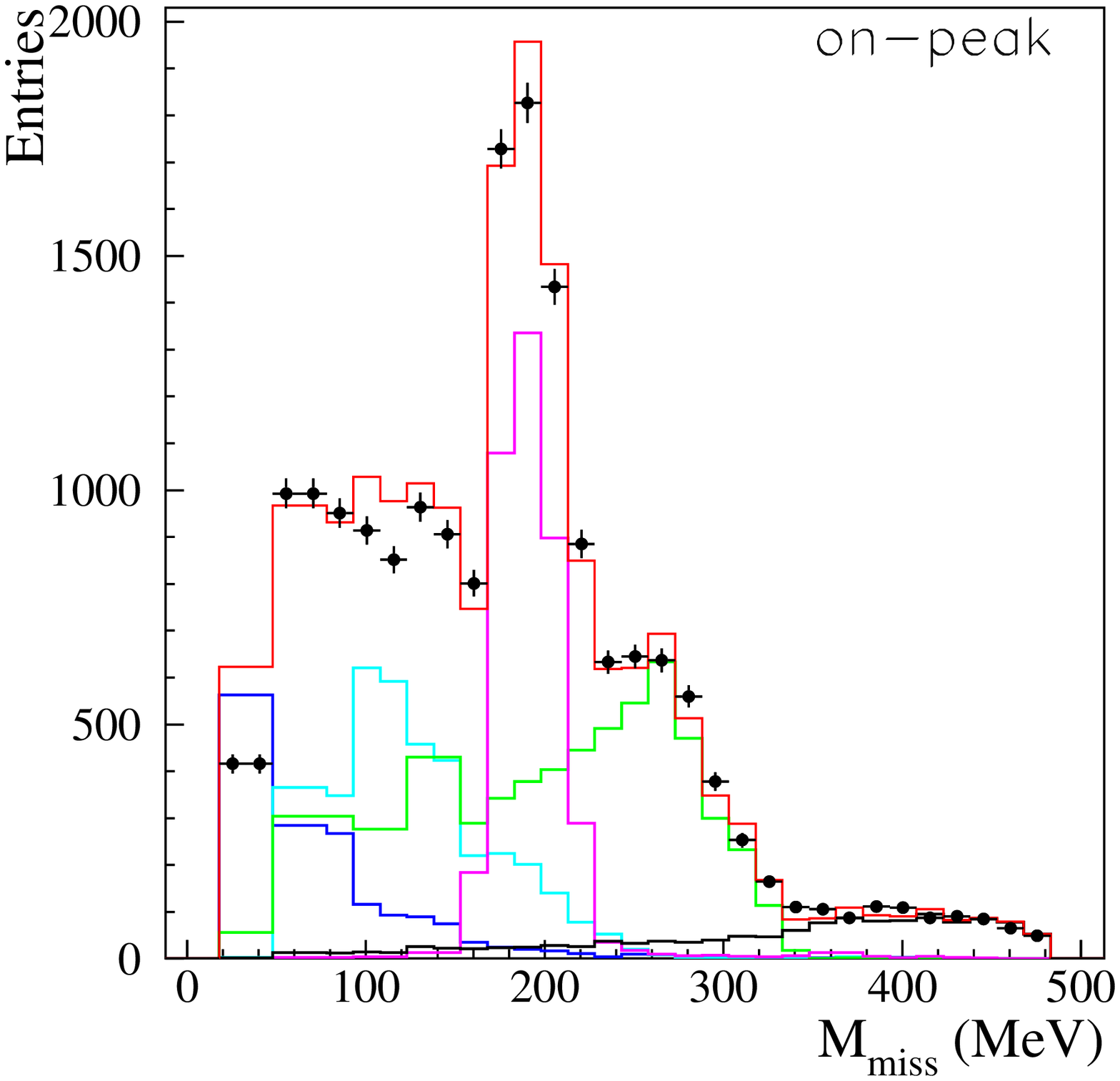}}
        {\includegraphics[scale=0.33]{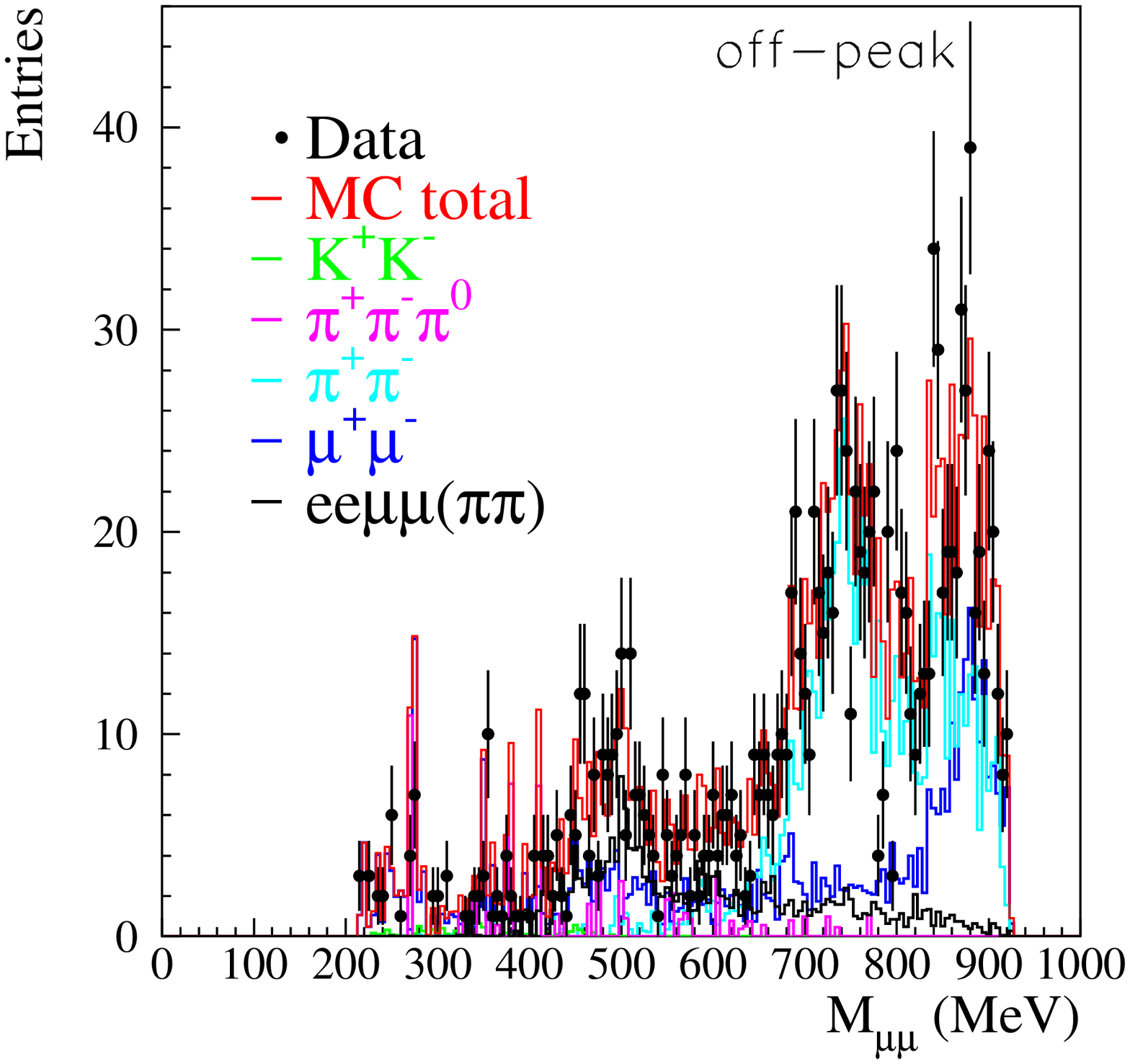}}\hspace{0.5cm}
        {\includegraphics[scale=0.33]{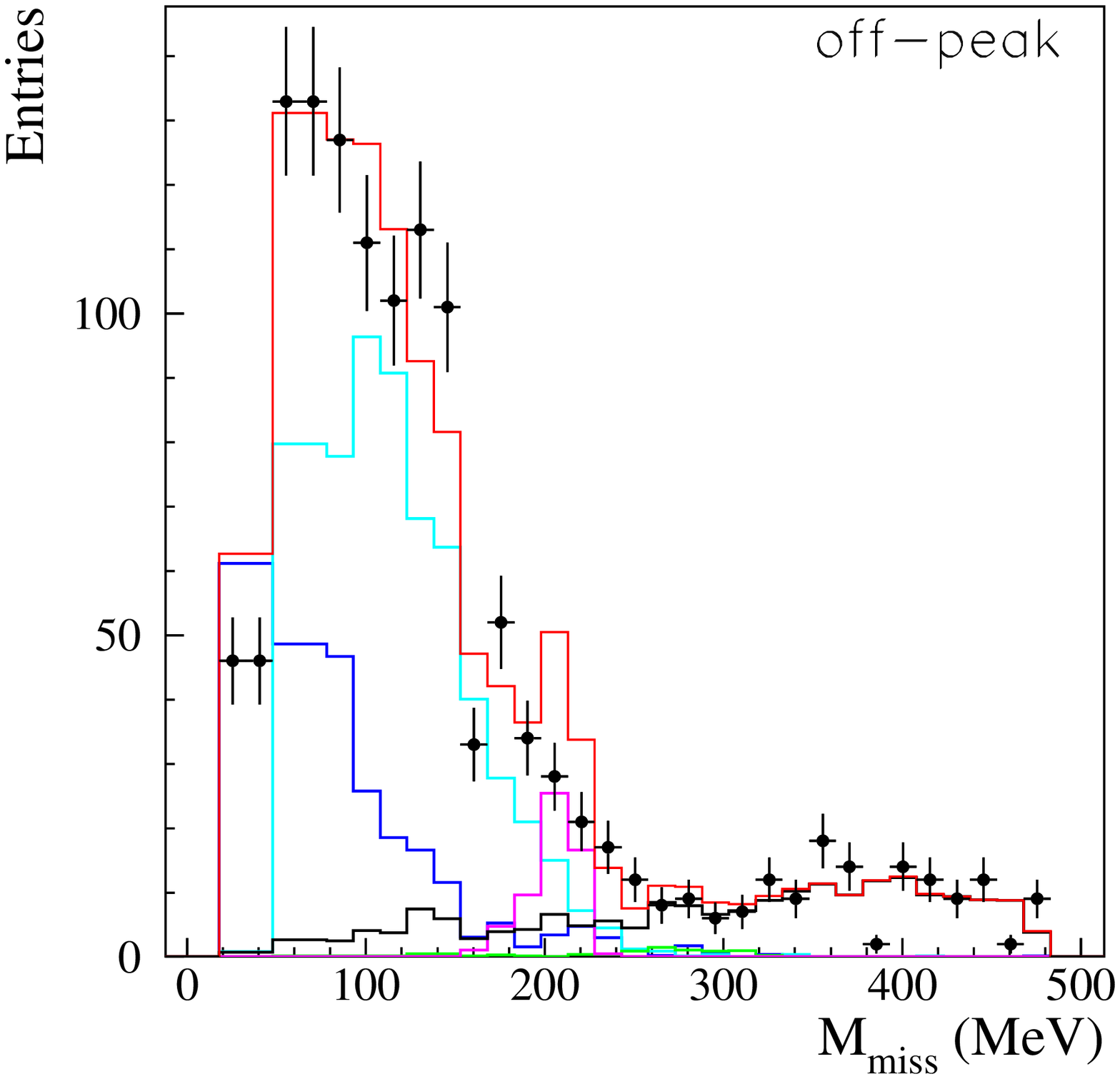}}
        \caption{Data - Monte Carlo comparison for the on-peak sample
(top plots) and off-peak sample (bottom plots).
Projections along the $M_{\mu \mu}$ axis (left plots); 
projections along the $M_{miss}$ axis (right plots).
Also shown are the various contributing backgrounds.}

\label{figdatamc}
    \end{center}
\end{figure}

Monte Carlo generators fully interfaced with the KLOE detector simulation 
program
were available for all the background processes but for the 
$e^+ e^- \rightarrow e^+ e^- \mu^+ \mu^-$ and 
$e^+ e^- \rightarrow e^+ e^- \pi^+ \pi^-$.
For these two processes the Courau generator program \cite{courau}
was used and the results 
smeared to keep into account the detector effects. 
 
As most of 
the signal is expected to populate a single bin of the mass 
distributions,
a 5$\times$5 bin matrix in the $M_{\mu \mu}$-$M_{miss}$ plane
was built and moved sliding all along the distributions of 
fig.\,\ref{figdati} 
both on data and Monte Carlo.
The presence of a possible signal was checked by using the central bin,
while the others were used for background evaluation. 
This was done by computing a data-Monte Carlo scale factor $k$
based on 
the sum of the contents of the 24 bins surrounding the central one 
in data ($DT_{24}$) and Monte Carlo ($MC_{24}$): 
$k={DT_{24} \over MC_{24}}$. The prediction for
the background in the central bin is then simply defined as 
the product of the central bin 
content in Monte Carlo rescaled by $k$.
The usage of the described scaling procedure 
allowed to reduce the systematic uncertainties 
due to the background evaluation (see sec. 5).  
Fig.\,\ref{figdatamc} 
shows the data-Monte Carlo 
comparison after the scaling correction for the 
on-peak and off-peak samples, projected along the $M_{\mu \mu}$ and $M_{miss}$
axes, together with the individual contributions of the different 
background processes. 
The agreement is satisfactory all over the populated regions of 
the distributions.

\section{Systematic errors}
Systematic uncertainties affect the signal efficiency evaluation
and the background estimate.
Several sources of systematic uncertainties 
in the signal efficiency 
evaluation from Monte Carlo were taken into account.

Uncertainties from the PID procedure were estimated by selecting 
samples of $e^+ e^- \rightarrow \mu^+ \mu^- \gamma$ in data and 
Monte Carlo, 
applying the
PID algorithms to one of the two tracks to increase the purity of the 
selection 
and studying on the opposite track the
data-Monte Carlo differences of the PID efficiency  
as a function of the track momentum. 
The total effect, 
defined as the 
average product of individual effects on the
single tracks,  was found to vary between 2\% and 3\%, depending on 
the boson masses.
A similar procedure was applied to evaluate the correction factors 
and systematic
uncertainties of the PID algorithms for pion identification, 
which affect the background evaluation.

The same  $e^+ e^- \rightarrow \mu^+ \mu^- \gamma$ samples selected in data
and in the simulation
were used to evaluate the effect of the cut on the vertex-IP distance.  
A correction to the Monte Carlo signal efficiencies of the order of 15\%, 
weakly dependent on $\cos\theta$, was derived and applied. An associated
systematic error of 0.5\% was estimated and added on the signal efficiency
evaluation.

The systematic uncertainty due to the usage of the EMC veto was evaluated 
by selecting samples of  
$\phi \rightarrow K^+ K^-$, $K^\pm \rightarrow \mu^\pm \nu$ 
in data and Monte Carlo. In this case, the cut on the vertex-IP distance
was slightly relaxed, in order to increase the size of the sample. 
A 2\% data-Monte Carlo difference was observed and
used both to correct the Monte Carlo efficiency and to quote a systematic
uncertainty due to this source.

The systematic uncertainty due to the kinematical preselections of the analysis
was estimated by varying track angles and momenta within their measurement 
errors by one standard deviation: a 1\% effect was ascribed
to this source.

The systematic uncertainty due to the binning choice was estimated by 
evaluating in the simulation the binomial statistical error on 
the fraction of the 
signal contained in one bin. This turned out to be of the order of 0.3\%, 
on average.

Finally, an average $\sim 1\%$ uncertainty was estimated due to the 
linear interpolation 
procedure in the signal efficiency evaluation process. 

The total 
systematic uncertainty on the signal efficiency was then evaluated as 
the quadratic sum of all the above effects. It never exceeded 4\%,  with an
average value of 3.5\%, very small when compared to the statistical 
uncertainties affecting this measurement.

Most of the systematic uncertainties in the background evaluation cancel 
in the scale factor ratio $k$. All the systematic
sources 
considered for the signal efficiency evaluation,
but those related to the linear interpolation procedure, 
were taken into account and their effect on the background estimate computed.

Additional effects were taken into account. The uncertainties on the
background process cross sections were varied
within their theoretical and measurement errors; a further 1\% 
uncertainty was
added for those related to the
photon-photon final states, for which no full simulation was available;
the uncertainty on the integrated luminosity was estimated to be 0.3\%.
 
The total systematic uncertainty 
on the background was evaluated as the quadratic sum of all the above 
effects.
It has an average value 
of 2.7\%  
with a very small tail extending up to 10\%.

\section{$p_0$ values and Upper Limits}
In order to evaluate the compatibility of the observed results with the
background only hypothesis ($p_0$ value)
and to derive upper limits 
to the parameters of the dark Higgstrahlung process,  
a Bayesian procedure
was set up. For each position of the $5 \times 5$ bin mass matrix, 
a likelihood function was devised based 
on uniform prior probabilities of the counting variables 
(constrained to be non negative) and on four Poissonian 
distributions representing the probabilities related respectively 
to the number of observed events in the central bin of the sliding 
matrix, the number of predicted background events in the same bin 
from Monte Carlo, the number of observed and predicted events in the
surronding 24 bins ($DT_{24}$ and  $MC_{24}$, entering 
in the scale factor ratio $k$). 
This procedure takes thus into full account the fluctuations due to the 
data and Monte Carlo statistics.
The systematic
uncertainties on the signal efficiency and on the background estimate 
were taken
into account by convolving the four Poissonian distributions with two 
correlated Gaussian distributions, with variances set equal to the
estimated systematic errors. Whenever the dark Higgsstrahlung process
was searched for, the small
fraction of signal expected outside the central bin of the 5x5 sliding matrix
was explicitly taken into account in the likelihood expression.

\begin{figure}[htp!]
    \begin{center}
        {\includegraphics[width=6.7cm]{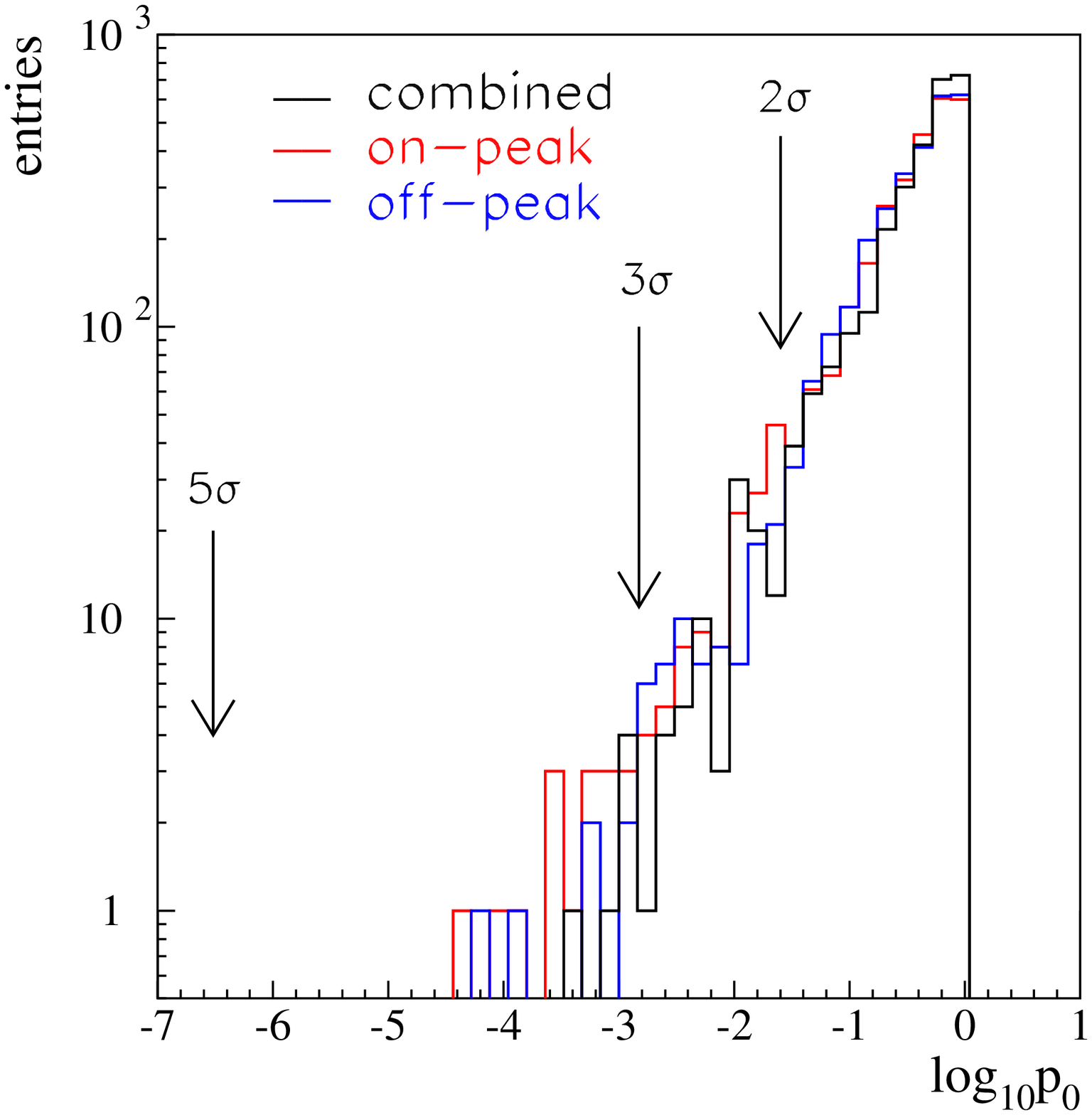}}\hspace{0.3cm}
        {\includegraphics[width=6.7cm]{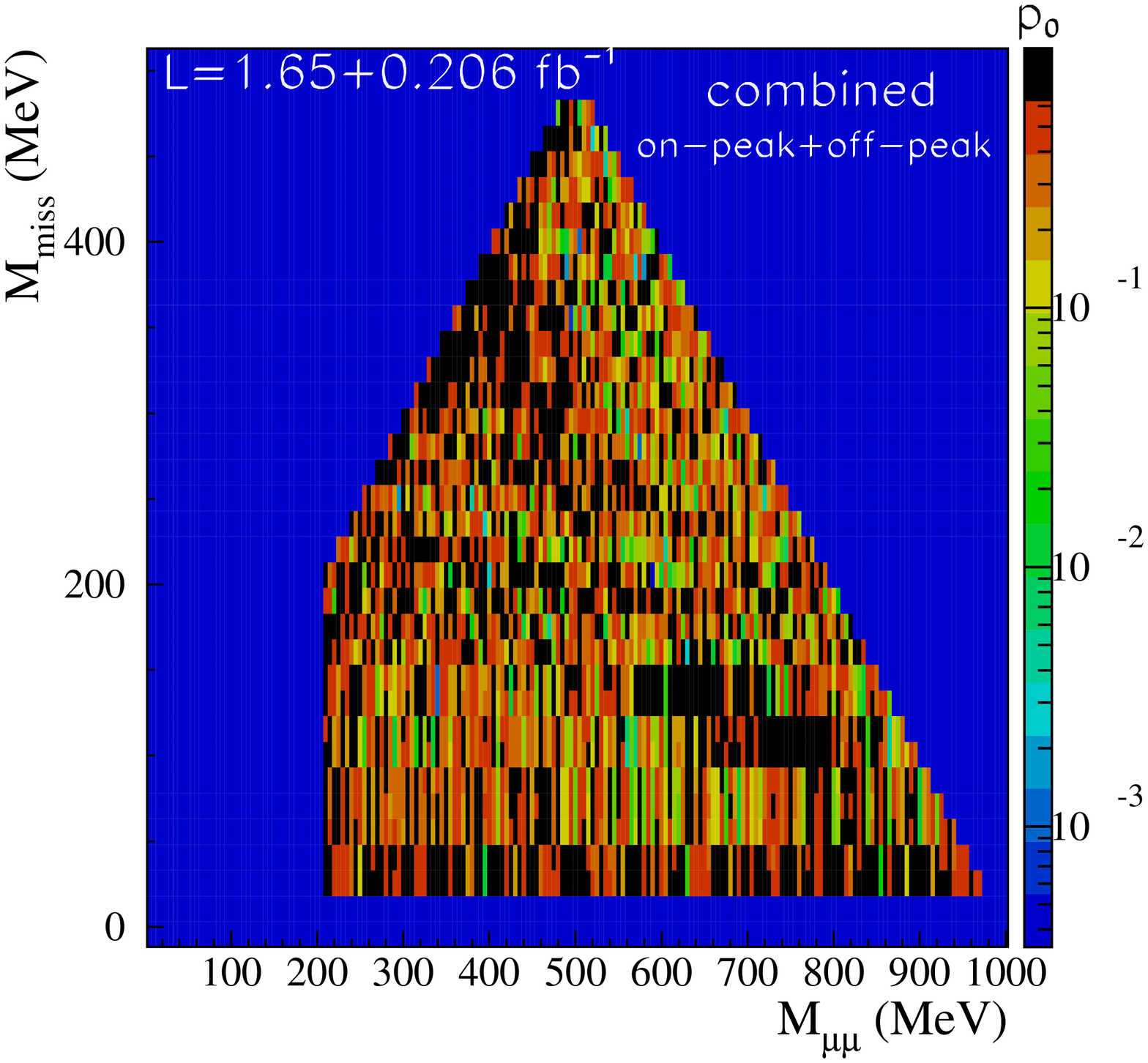}}
        \caption{Left: $p_0$ value distribution for the on-peak sample 
(red line),
off-peak sample (blue line), combined sample (black line).
Right: $p_0$ values for the combined result.}
\label{figpval1d}
    \end{center}
\end{figure}

The $p_0$ distributions
for the on-peak, off-peak  and combined samples are shown in 
fig.\,\ref{figpval1d}, left plot. 
Fig.\,\ref{figpval1d}, right plot,
shows the computed $p_0$ values as a function of
$M_{\mu \mu}$-$M_{miss}$ masses for the 
combined sample. 
There are
three values exceeding the threshold corresponding to a 3$\sigma$ excess, 
while 4.2 were expected
on probabilistic base. 
The excess significance of those points 
(see fig.\,\ref{figpval1d}) are at the level of 3.1$\sigma$,
3.2$\sigma$ and 3.4$\sigma$.
In the on-peak and off-peak samples the most significant values exceeding
the 3$\sigma$ threshold are at the level of 3.9$\sigma$ and 3.8$\sigma$
respectively (see fig.\,\ref{figpval1d}, left plot)
These excesses, even though
at quite interesting level, are then lost in the combination of the two samples,
becoming fluctuations of average size.

\begin{figure}[htp!]
    \begin{center}
        {\includegraphics[width=6.7cm]{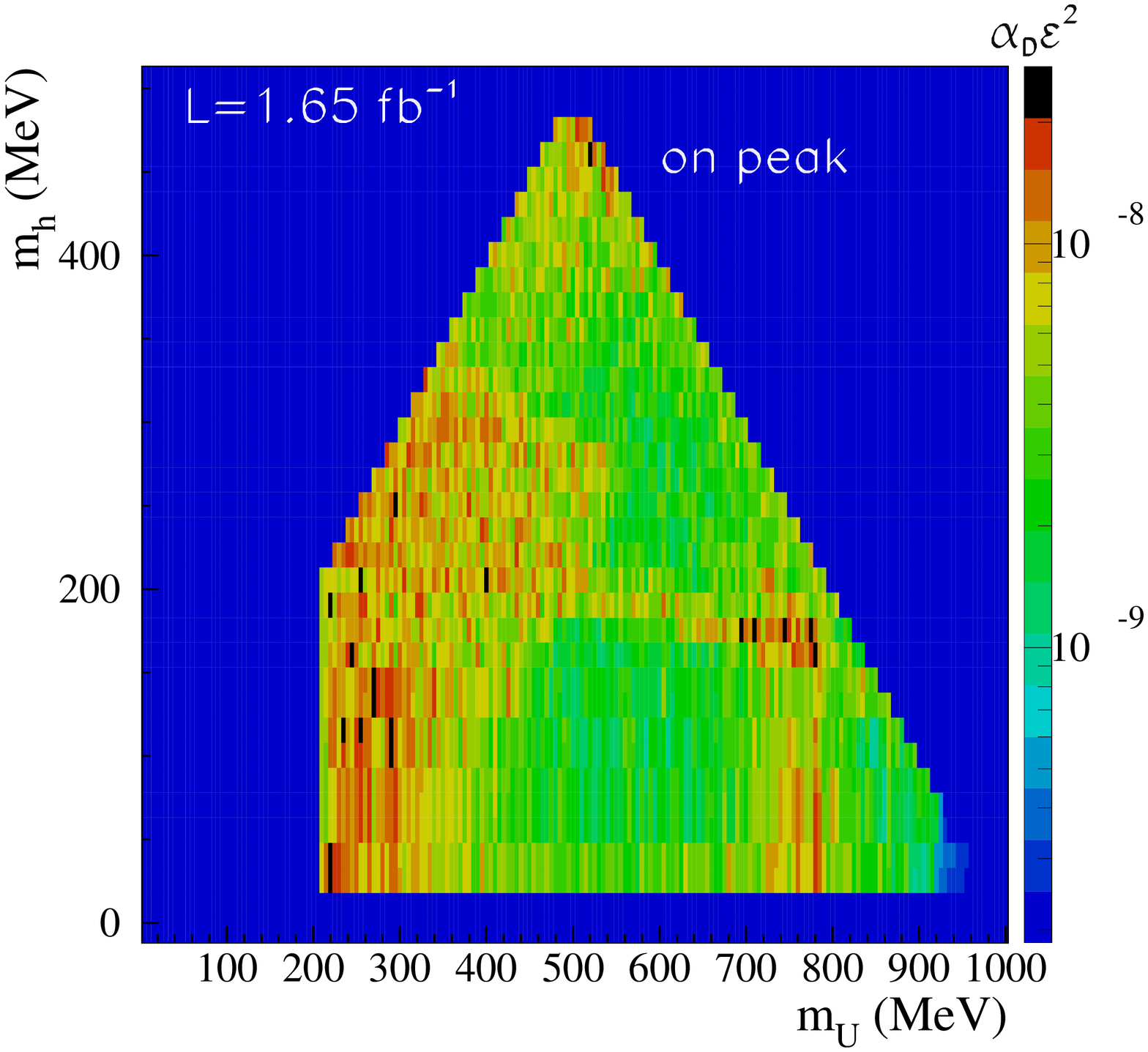}}\hspace{0.3cm}
        {\includegraphics[width=6.7cm]{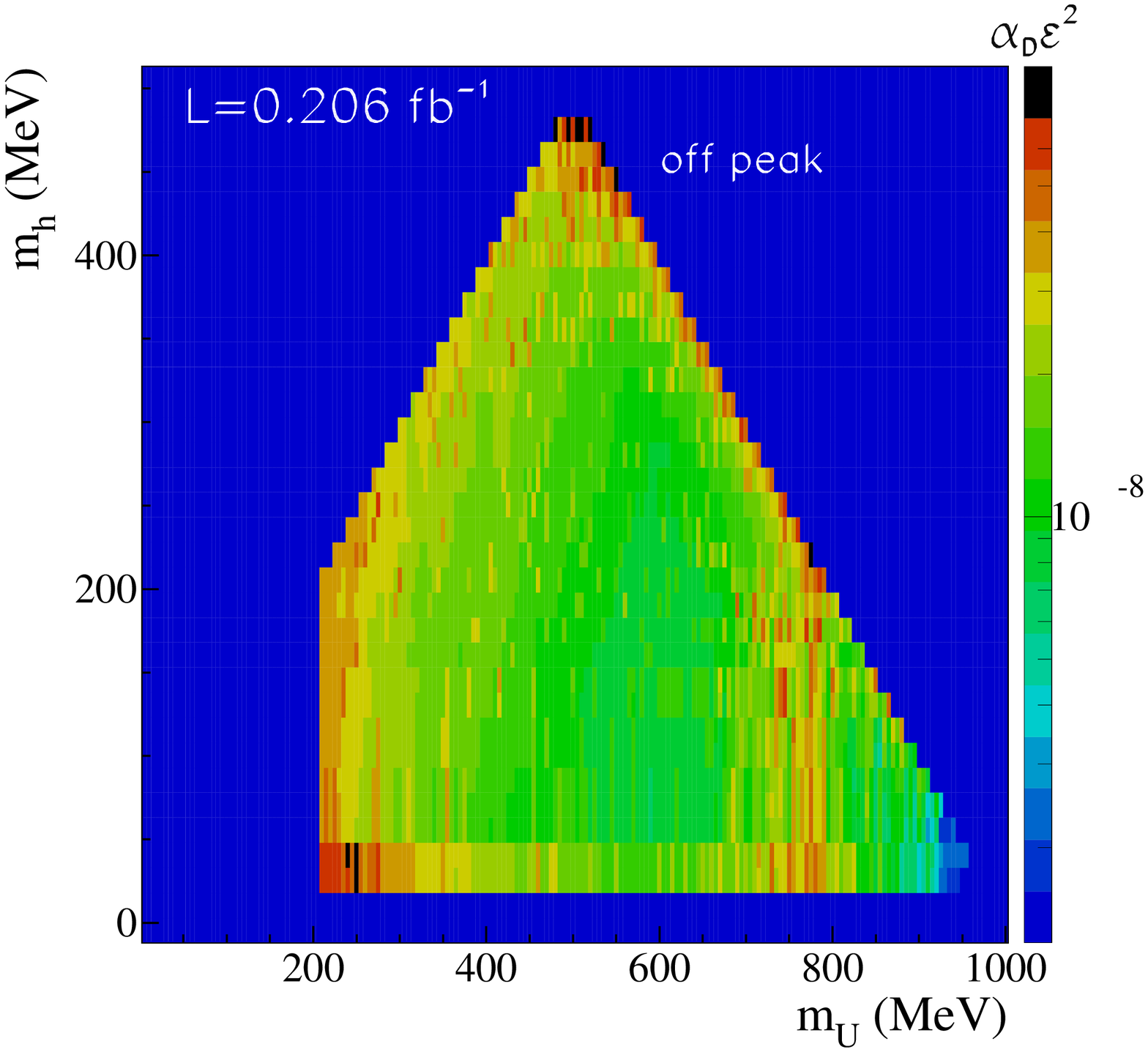}}
        \caption{90\% CL upper limits in $\alpha_D \times \epsilon^2$ for 
the on-peak sample (left plot) and off-peak sample (right plot).}
\label{figlim2d}
    \end{center}
\end{figure}

\begin{figure}[htp!]
    \begin{center}
        {\includegraphics[scale=0.33]{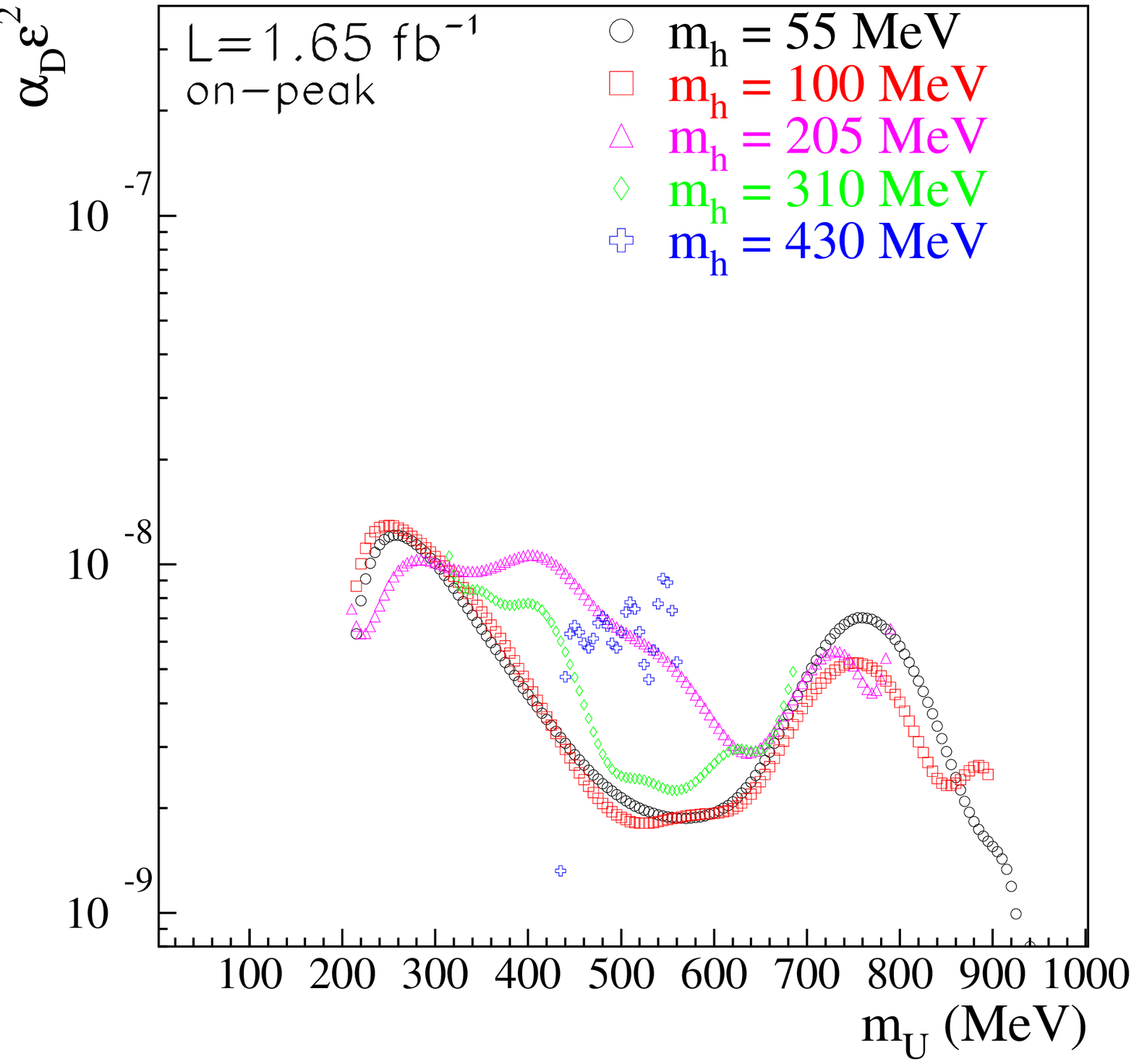}}\hspace{0.5cm}
        {\includegraphics[scale=0.33]{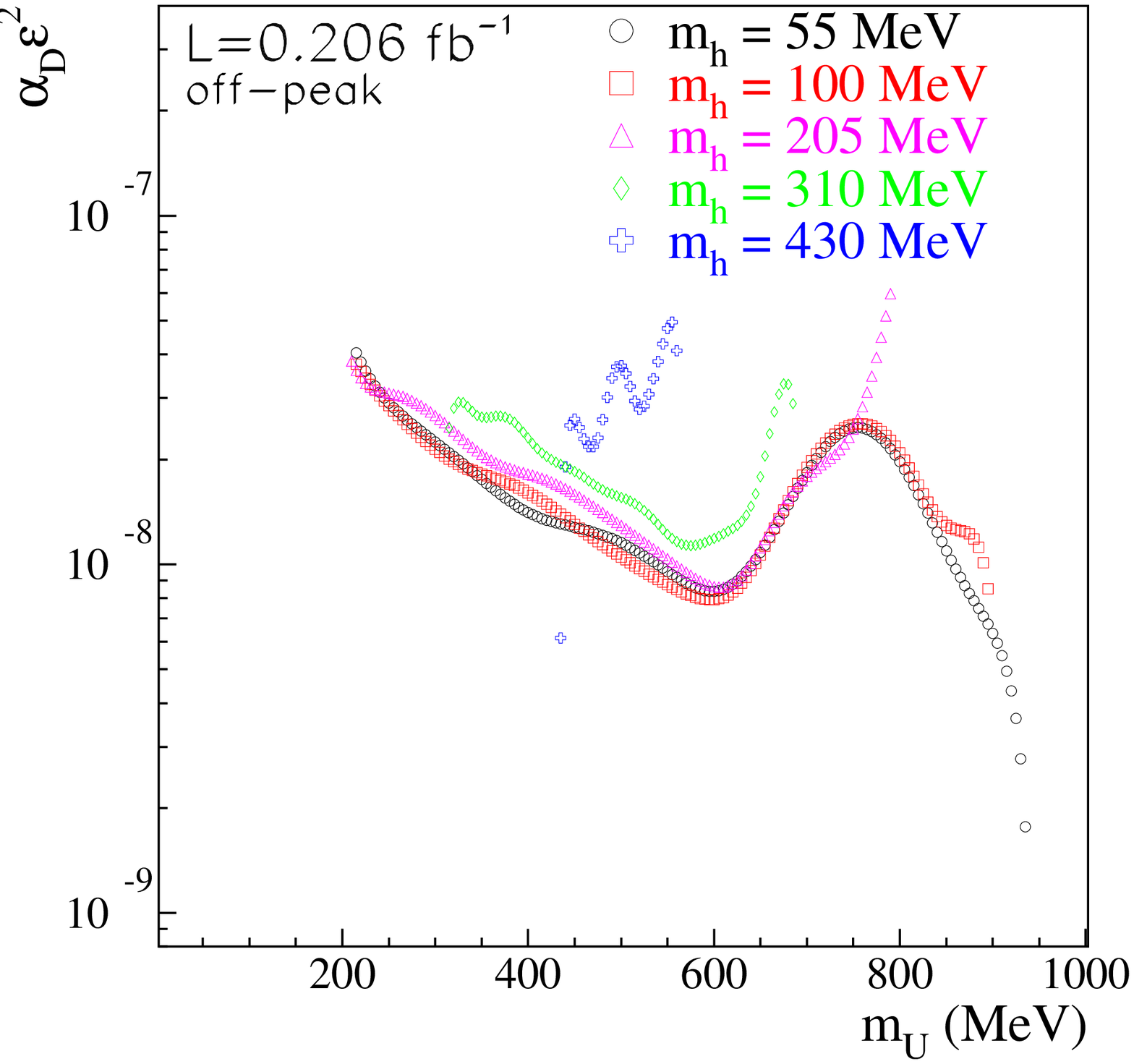}}
        {\includegraphics[scale=0.33]{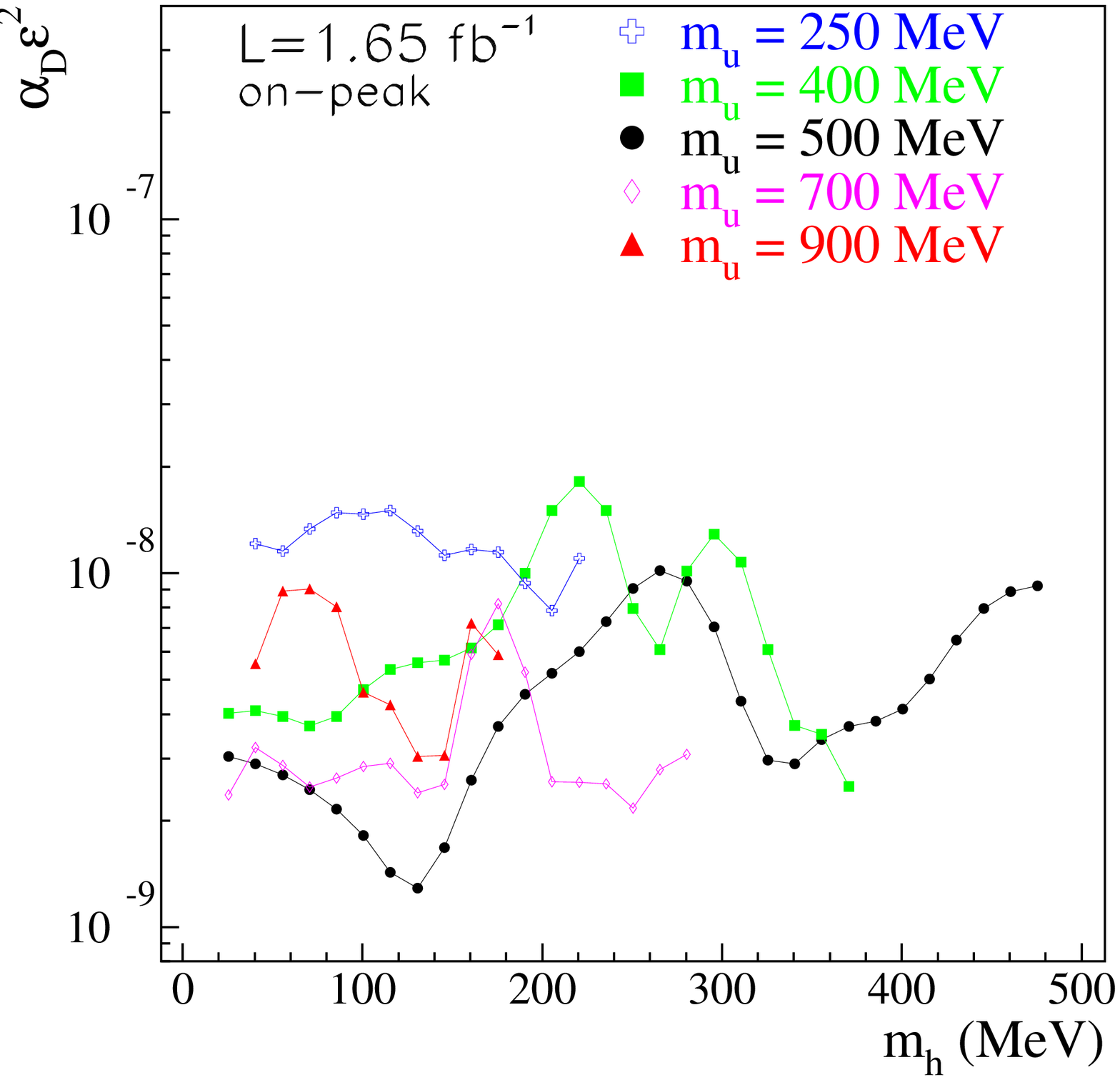}}\hspace{0.5cm}
        {\includegraphics[scale=0.33]{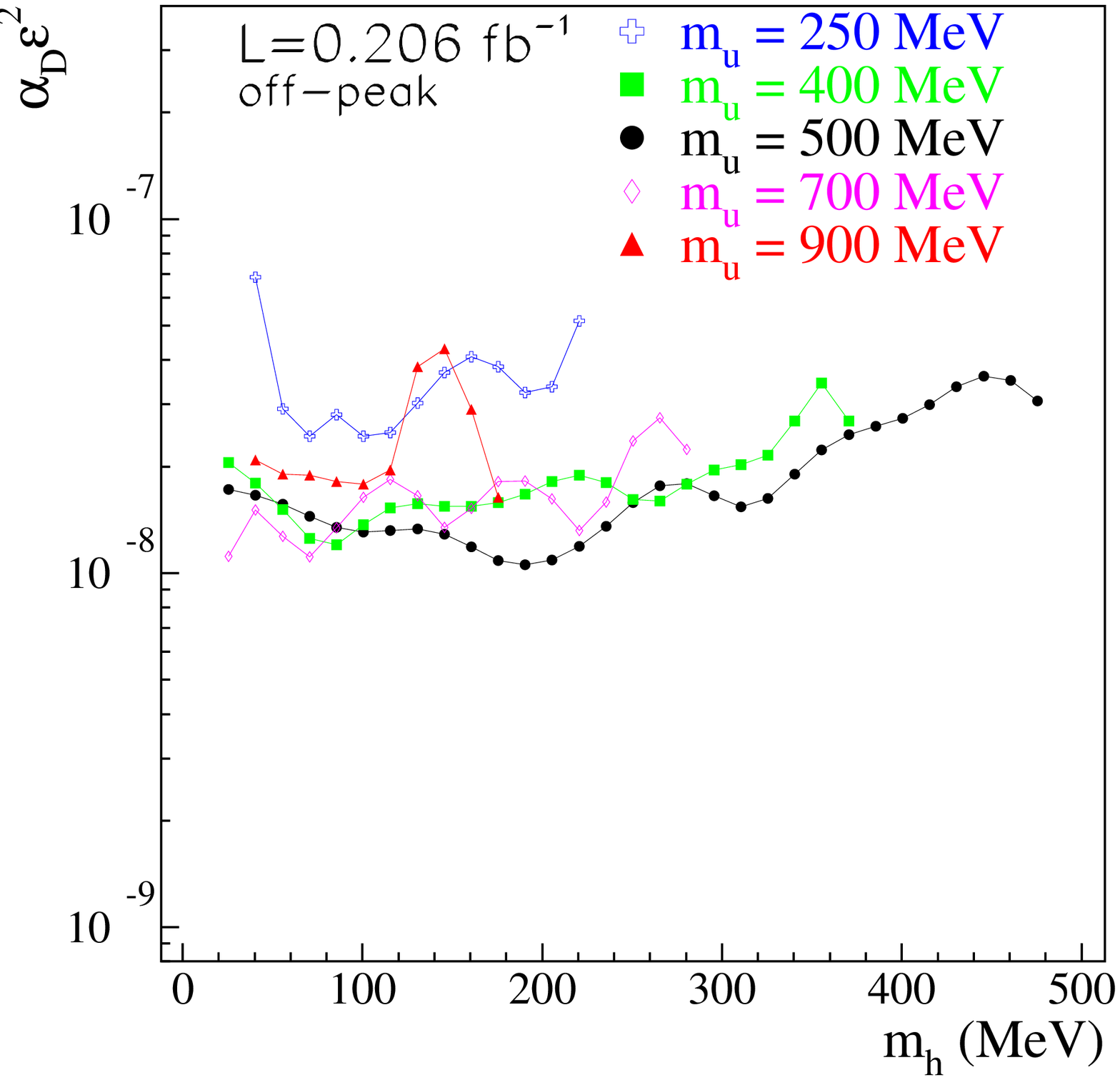}}
        \caption{Top plots: 90\% CL upper limits 
in $\alpha_D \times \epsilon^2$ as a function of 
$m_U$ for different values of $m_{h^\prime}$  
for the on-peak sample (top, left)
and off-peak sample (top, right). Bottom plots: same limits
as a function of $m_{h^\prime}$  for different 
values of $m_U$ for the on-peak sample (bottom, left) and off-peak sample 
(bottom, right). }
\label{figlimmu}
    \end{center}
\end{figure}

\begin{figure}[htp!]
    \begin{center}
        {\includegraphics[width=6.7cm]{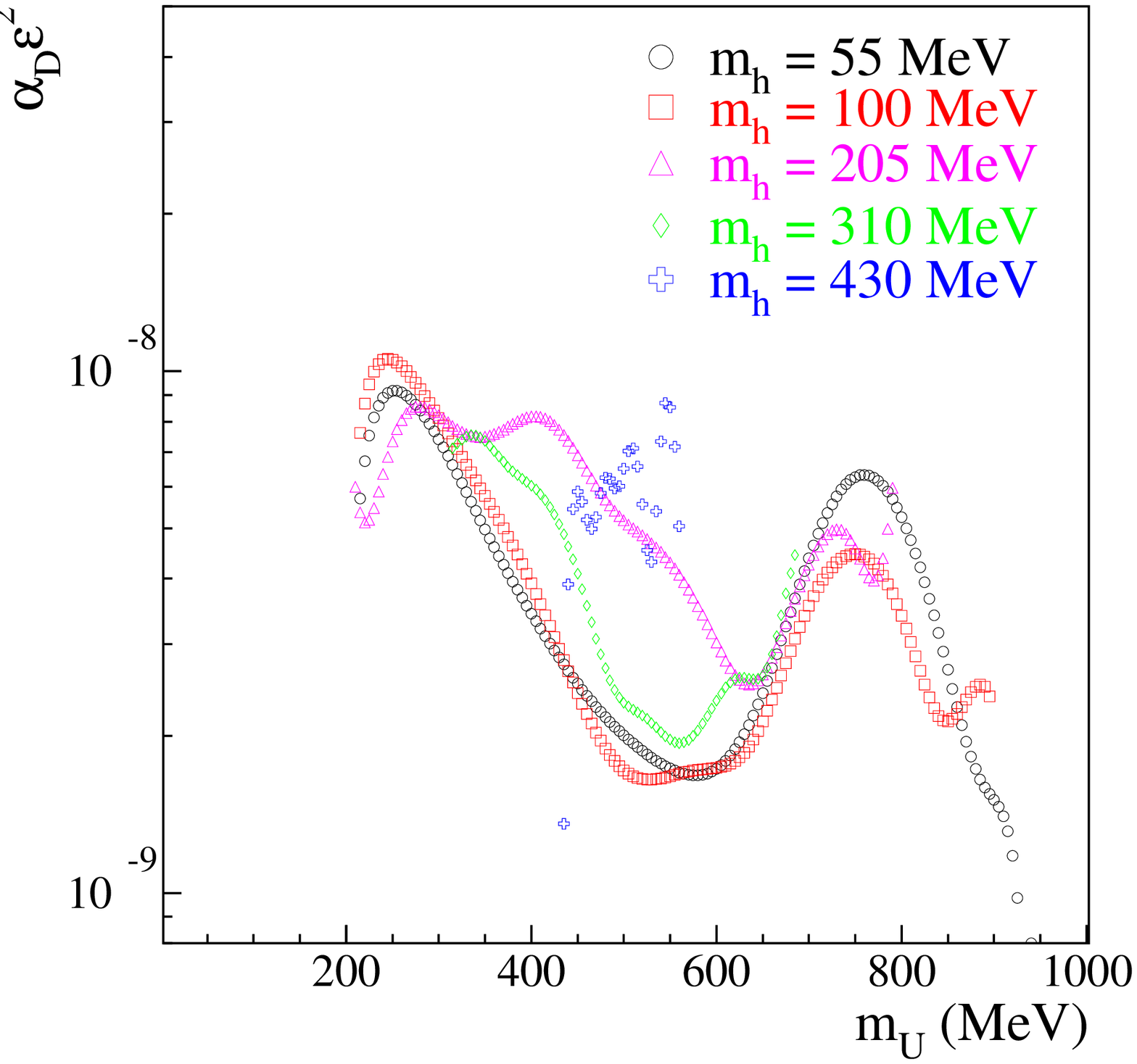}}\hspace{0.3cm}
        {\includegraphics[width=6.7cm]{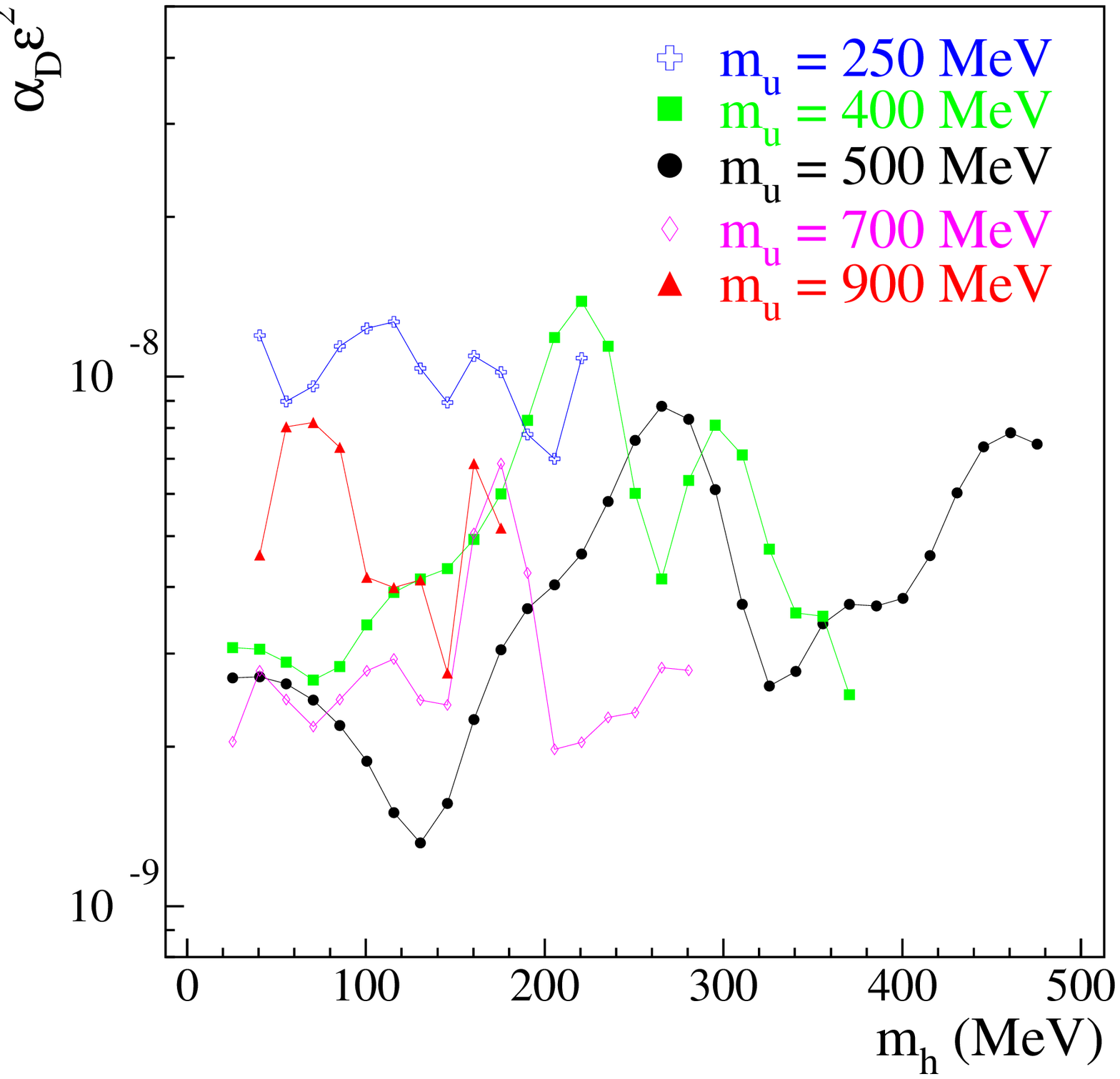}}
        \caption{Combined 90\% CL upper limits 
in $\alpha_D \times \epsilon^2$ as a function of 
$m_U$ for different values of $m_{h^\prime}$ 
(top plot) and as a function of $m_{h^\prime}$ for different 
values of $m_U$ (bottom plot).}
\label{figlim1d}
    \end{center}
\end{figure}

As no evidence of the dark Higgsstrahlung process was found, 90\%
confidence level Bayesian upper limits on the number of events  were 
derived bin by bin
in the $M_{\mu \mu}$-$M_{miss}$ plane, 
separately for the on-peak and off-peak samples, and then converted in
terms of $\alpha_D \times \epsilon^2$.
They are shown in fig.\,\ref{figlim2d}. 
Fig.\,\ref{figlimmu} 
shows the
on-peak and off-peak 90\% CL upper limits projected along the 
 $m_U$ and $m_{h^\prime}$ axes
after a slight smoothing to make them more readable.
The different curves in 
$m_U$ ($m_{h^\prime}$)
correspond to different values of $m_{h^\prime}$ ($m_U$).
These results were then combined
by taking into account the different integrated luminosities 
of the two samples and the respective
signal
efficiencies and cross sections. 
The combined results are almost everywhere dominated
by the on-peak sample, because of the larger available statistics,
with the exception of some very noisy background regions.  
They are shown in fig.\,\ref{figlim1d}. 
These limits are largely dominated by the data statistics.
Values as low as 10$^{-9}\div$10$^{-8}$ 
of the product $\alpha_D \times \epsilon^2$
are excluded at 90\% CL for a large range of the dark photon 
and dark Higgs masses.

\section{Conclusions}
\label{conclusions}

A search for the dark Higgsstrahlung process 
\hu, $U \rightarrow \mu^+ \mu^-$, $h^\prime$ invisible,
has been performed by KLOE
in the range 
$2 m_\mu < m_U <$ 1000 MeV with $m_{h^\prime} < m_U$. No evidence for signal has
been observed and upper limits on the product of the kinetic mixing 
parameter $\epsilon$
and the dark coupling constant $\alpha_D$ have been set in the range  
10$^{-9}\div$10$^{-8}$ in $\alpha_D \times \epsilon^2$.
With the arbitrary hypothesis $\alpha_D = \alpha_{em}$ these measurements 
translate into limits on
the kinetic mixing parameter $\epsilon$ in the range  10$^{-4}\div$10$^{-3}$.

\section{Acknowledgments} 
We warmly thank our former KLOE colleagues for the access to the data collected during the KLOE data taking campaign.
We thank the DA$\Phi$NE team for their efforts in maintaining low background running conditions and their collaboration during all data taking. We want to thank our technical staff: 
G.F. Fortugno and F. Sborzacchi for their dedication in ensuring efficient operation of the KLOE computing facilities; 
M. Anelli for his continuous attention to the gas system and detector safety; 
A. Balla, M. Gatta, G. Corradi and G. Papalino for electronics maintenance; 
M. Santoni, G. Paoluzzi and R. Rosellini for general detector support; 
C. Piscitelli for his help during major maintenance periods. 
This work was supported in part by the EU Integrated Infrastructure Initiative Hadron Physics Project under contract number RII3-CT- 2004-506078; by the European Commission under the 7th Framework Programme through the `Research Infrastructures' action of the `Capacities' Programme, Call: FP7-INFRASTRUCTURES-2008-1, Grant Agreement No. 227431; by the Polish National Science Centre through the Grants No. 0469/B/H03/2009/37, 0309/B/H03/2011/40, DEC-2011/03/N/ST2/02641, \\
2011/01/D/ST2/00748, 2011/03/N/ST2/02652, 2013/08/M/ST2/00323 and by the Foundation for Polish Science through the MPD programme and the project HOMING PLUS BIS/2011-4/3.




\end{document}